\documentclass[letterpaper,twocolumn,10pt]{article}
\usepackage{usenix-2020-09}

\usepackage{array}
\usepackage[framemethod=TikZ]{mdframed}
\usepackage{xcolor}
\usepackage{tikz}
\usepackage{amsmath}
\usepackage{filecontents} %
\usepackage[normalem]{ulem}
\usepackage{amssymb}
\usepackage{multirow}
\usepackage{wasysym}
\usepackage{hyperref}
\hypersetup{
	unicode=true,
        bookmarksnumbered=true,
	colorlinks=true,
	linkcolor={red!70!black},
	citecolor={red!70!black},
	urlcolor={blue!70!black},
	pdfborder={0 0 0}
}
\usepackage[capitalize,nameinlink]{cleveref}
\usepackage[utf8]{inputenc}
\usepackage{amssymb}
\usepackage{array}
\usepackage{longtable}
\usepackage{xspace}
\usepackage{booktabs}
\usepackage{flushend}
\usepackage{makecell}
\usepackage{boldline}
\usepackage{boxedminipage}
\usepackage{booktabs}
\usepackage{enumitem}
\usepackage{cite}
\usepackage{comment}

\newcommand{\myparagraph}[1]{\vspace{\smallskipamount}\noindent\textbf{#1.\xspace}}
\newcommand{\myparagraphemph}[1]{\vspace{\smallskipamount}\noindent\emph{#1.\xspace}}
\newcommand{\eg}{\emph{e.g.}\xspace}

\newcommand{\ie}{\emph{i.e.}\xspace}

\newtoggle{showmarks}

\newcommand{\CompanyX}{ProviderX\xspace}
\newcommand{\CosmosDB}{CloudDB\xspace}

\iftoggle{showmarks}{
  \newcommand\lexiang[1]{\textcolor{magenta}{Lexiang: #1}}
  \newcommand\chetan[1]{\textcolor{cyan}{Chetan: #1}}
  \newcommand\anjaly[1]{\textcolor{cyan}{Anjaly: #1}}
  \newcommand\inigo[1]{\textcolor{red}{IG: #1}}
  \newcommand\ricardo[1]{\textcolor{yellow}{Ricardo: #1}}
  \newcommand\raphael[1]{\textcolor{purple}{Raphael: #1}}
  \newcommand\timmy[1]{\textcolor{olive}{Timmy: #1}}
  \newcommand\minghua[1]{\textcolor{green!55!blue}{Minghua: #1}}
  \newcommand\jue[1]{\textcolor{red!55!blue}{Jue: #1}}
  \newcommand\xiaoting[1]{\textcolor{pink}{Xiaoting: #1}}
  \newcommand\TODO[1]{\textcolor{blue}{TODO: #1}}
  \newcommand\pulkit[1]{\textcolor{blue}{Pulkit: #1}}
  \newcommand\pantea[1]{\textcolor{teal}{PZ: #1}}
  \newcommand\rodrigo[1]{\textcolor{orange}{RF: #1}}
  \newcommand\review[1]{\textcolor{green}{Review: #1}}
  \newcommand\edit[1]{\textcolor{orange}{#1}}
}{
  \newcommand\lexiang[1]{\unskip}
  \newcommand\chetan[1]{\unskip}
  \newcommand\anjaly[1]{\unskip}
  \newcommand\inigo[1]{\unskip}
  \newcommand\ricardo[1]{\unskip}
  \newcommand\raphael[1]{\unskip}
  \newcommand\timmy[1]{\unskip}
  \newcommand\minghua[1]{\unskip}
  \newcommand\jue[1]{\unskip}
  \newcommand\xiaoting[1]{\unskip}
  \newcommand\TODO[1]{\unskip}
  \newcommand\pulkit[1]{\unskip}
  \newcommand\pantea[1]{\unskip}
  \newcommand\rodrigo[1]{\unskip}
  \newcommand\review[1]{\unskip}
  \newcommand\edit[1]{#1}
}

\usepackage[compact]{titlesec}

\begin{document}

\date{}

\title{Workload Intelligence: Punching Holes Through the Cloud Abstraction}

\author{
Lexiang Huang\textsuperscript{1,3},
Anjaly Parayil\textsuperscript{3},
Jue Zhang\textsuperscript{3},
Xiaoting Qin\textsuperscript{3},
Chetan Bansal\textsuperscript{3},
\\
Jovan Stojkovic\textsuperscript{2,3},
Pantea Zardoshti\textsuperscript{3},
Pulkit Misra\textsuperscript{3},
Eli Cortez\textsuperscript{3},
Raphael Ghelman\textsuperscript{3},
\'I\~nigo Goiri\textsuperscript{3},
\\
Saravan Rajmohan\textsuperscript{3},
Jim Kleewein\textsuperscript{3},
Rodrigo Fonseca\textsuperscript{3},
Timothy Zhu\textsuperscript{1},
Ricardo Bianchini\textsuperscript{3}
\vspace{.2in}
\\
$^1$The Pennsylvania State University
\qquad 
$^2$University of Illinois at Urbana Champaign
\qquad 
$^3$Microsoft
}

\maketitle

\begin{abstract}
Today, cloud workloads are essentially opaque to the cloud platform. Typically, the only information the platform receives is the virtual machine (VM) type and possibly a decoration to the type (\eg, the VM is evictable).
Similarly, workloads receive little to no information from the platform; generally, workloads might receive telemetry from their VMs or exceptional signals (\eg, shortly before a VM is evicted).
\edit{
The narrow interface between workloads and platforms has several drawbacks:
(1) a surge in VM types and decorations in public cloud platforms complicates customer selection;
(2) essential workload characteristics (\eg, low availability requirements, high latency tolerance) are often unspecified, hindering platform customization for optimized resource usage and cost savings;
and (3) workloads may be unaware of potential optimizations or lack sufficient time to react to platform events.
}

In this paper, we propose a framework, called Workload Intelligence (WI), for dynamic bi-directional communication between cloud workloads and cloud platform.
Via WI, workloads can programmatically adjust their key characteristics, requirements, and even dynamically adapt behaviors like VM priorities.
In the other direction, WI allows the platform to programmatically inform workloads about upcoming events, opportunities for optimization, among other scenarios.
Because of WI, the cloud platform can drastically simplify its offerings, reduce its costs without fear of violating any workload requirements, and reduce prices to its customers on average by 48.8\%.
\end{abstract}

\section{Introduction}

\footnotetext[1]{Lexiang Huang was an intern at Microsoft.}
\footnotetext[2]{Jovan Stojkovic was an intern at Microsoft.}

From its inception with a simple interface of offering virtual machines to users, today's cloud has grown incredibly complex for both cloud providers and workload owners. Cloud providers constantly seek to improve their efficiency, which gives rise to many optimizations. 
Some of these are exposed to workload owners via VM types (\eg, Spot VMs~\cite{ec2-spot,azure-spot,google-spot}, Harvest VMs~\cite{ambati2020slo}, Burstable VMs~\cite{wang2017using,ec2-burstable,azure-burstable}) and dedicated interfaces (\eg, auto-scaling~\cite{mao2011auto}), whereas other optimizations are internal to the cloud (\eg, oversubscription~\cite{householder2014cloud}, pre-provisioning~\cite{yao2021pps}).
These can reduce costs, and/or improve efficiency, performance, sustainability, or reliability.
Even with these interfaces, cloud workloads are essentially opaque since cloud platforms lack visibility into the desires and expectations of workload owners.
Hence, the platforms are limited to inferring workload characteristics, adding/extending interfaces, or being overly conservative.
This results in undesirable performance effects, increased complexity, and higher costs, negatively impacting both workload owners and providers.

On the other hand, workload owners lack visibility and control over the impact of the provider’s optimizations on their workloads.
Typically, they only receive telemetry from their VMs and any exceptional signals (\eg, shortly before a VM is evicted).
As a result, performance-conscious workload owners need to develop workarounds to cope with the variability and unknowns (\eg, spawning multiple VMs to find ones without noisy neighbors~\cite{lloyd2017mitigating}).

\inigo{The following paragraph is repeated from the abstract.}

\myparagraph{Revisiting the cloud interface}
The narrow communication interface between workloads and platform has multiple negative effects:
(1) the number of VM types and decorations has exploded in public cloud platforms, making it difficult for workload owners to select the ideal ones;
(2) many important workload characteristics (\eg, low availability requirements, high tolerance to latency) are never made explicit, so the platform is unable to customize its service to them (\eg, by optimizing their resource usage and passing any dollar savings to workload owners);
and (3) workloads often are unaware of optimizations that they could make or do not have enough time to react to platform events.

\begin{table*}[t!]
    \centering
    \scriptsize{
    \begin{tabular}{cc@{}c@{}c@{}@{}@{}@{}|}
    \toprule
    \textbf{Category} & \textbf{Workload characteristics} & \textbf{Core usage} \\
    \midrule
    \multirow{3}{*}{Scalability} & Stateless nature &
    \begin{tabular}{ccc}
        Stateless  & Partially stateless & Stateful\\
        \midrule
        45.5\% & 17.4\% & 37.1\% \\
    \end{tabular}
    \tabularnewline
    \cline{2-3}
    & Deployment time requirements &
    \begin{tabular}{cc}
        Strict  & Not strict\\
        \midrule
        28.5\% & 71.5\% \\
    \end{tabular}
    \tabularnewline
    \midrule
    \multirow{3}{*}{Reliability} & Availability  & 
    \begin{tabular}{cccccc}
        Five Nines  & Four Nines & Three Nines & Two Nines & One Nine & None \\
        \midrule
        2.4\% & 34.5\% & 58.0\% & 3.9\% & 0.5\% & 0.4\% \\
    \end{tabular}
    \tabularnewline
    \cline{2-3}
    & Preemptibility  &
    \begin{tabular}{ccccccc}
        0\% & 0-20\% & 20-40\% & 40-60\% &60-80\% & 80-100\% & 100\% \\
        \midrule
        39.3\% & 41.1\% & 4.8\% & 6.5\% & 0.3\% & 1.8\% & 6.1\% \\
    \end{tabular}
    \tabularnewline
    \cline{1-3}
    Performance & Delay tolerance &
    \begin{tabular}{cc}
        Delay tolerant  & Delay sensitive\\
        \hline
        24.5\% & 75.5\% \\
    \end{tabular}
    \tabularnewline
    \midrule
    Geographical & Region independence &
    \begin{tabular}{ccc}
       Region-agnostic & Partially region-agnostic & Not region-agnostic\\
       \midrule
       47.5\% &
       13.9\% & 
       38.6\%
    \end{tabular}
    \tabularnewline
    \bottomrule
    \end{tabular}
    }
    \caption{
    Overview of cloud workload characteristics and their core usage based on an internal survey.
    }
    \label{tab:survey-results}
\end{table*}

\myparagraph{Our work}
\edit{
In this paper, we study how internal workloads use cloud platform optimizations and identify the fundamental characteristics that cloud platform optimizations require to operate.
Based on this characterization, we propose a framework, called Workload Intelligence (WI), for dynamic bi-directional communication between cloud workloads and cloud platform.
}
WI enables workloads to programmatically communicate their key characteristics, requirements, dynamic changes, and shifting behaviors.

In the other direction, WI allows the platform to programmatically notify workloads about upcoming events, optimization opportunities, and other useful scenarios.
With WI, the cloud platform can drastically simplify its offerings, reduce costs without fear of violating any workload requirements, and lower prices for workload owners.

In building WI, we address three key challenges:
(1) creating a general, extensible, and incrementally adoptable interface for workloads to express their main characteristics, requirements, and dynamic behaviors;
(2) developing a dynamic framework for the platform to seamlessly interact with workloads, even at a potentially high rate, while preventing potential attacks or bugs that could harm it or other workloads;
and (3) enabling the platform to process received information intelligently, maximizing optimization opportunities, and maintaining quality of service.

We start by exploring the characteristics and requirements of 188 real cloud workloads at a major cloud provider (\CompanyX for blind review).
Next, we discuss optimizations that the platform can perform by knowing these characteristics and requirements, enabled by its bi-directional communication capability with workloads.
Afterward, we introduce the design and implementation of WI, focusing mainly on how WI effectively tackles the four challenges.
To evaluate WI, we explore various workload and optimization scenarios.

We conclude that it is possible to build a frameworks for bi-directional communication between workloads and the platform in a safe and effective manner.
By punching holes in the cloud abstraction for this communication, WI can simplify cloud offerings and make platforms more efficient and cost-effective, while providing excellent service to workloads.

\myparagraph{Summary}
We make the following contributions:
\begin{itemize}[nosep,leftmargin=*]
    \item We identify the fundamental workload characteristics that cloud platform optimizations require to operate and show their savings potential.
    \item We develop Workload Intelligence (WI), a novel and extensible framework that enables bidirectional communication between workloads and the cloud provider for improving cloud efficiency.
    \item We evaluate the applicability and potential of WI across ten cloud optimizations at \CompanyX and demonstrate that WI can on average save workload costs by 48.8\%.
\end{itemize}

\section{Characterizing workloads in the cloud}
\label{sec:characterization}
\begin{table*}[t!]
    \centering
    \scriptsize{
    \begin{tabular}{c|ccccc}
         \toprule
         \textbf{Cloud} &
         \textbf{Cloud} &
         \textbf{User Benefit} &
         \textbf{Min} &
         \textbf{Max} &
         \textbf{Platform Benefit}
         \\
         \textbf{Optimization} &
         \textbf{Resources} &
         \textbf{(Average)} & %
         \textbf{Pricing} &
         \textbf{Pricing} &
         \textbf{Model} %
         \\
         \midrule
         Auto-scaling &
         Compute &
         19\% less cost, $\downarrow$ Carbon & %
         \multicolumn{2}{c}{Average number of regular VMs} &
         Compute allocation
         \\
         Spot VMs &
         Spare compute &
         85\% less cost & %
         \multicolumn{2}{c}{15\% regular VM} &
         Compute allocation
         \\
         Harvest VMs &
         Spare compute &
         91\% less cost & %
         Spot VM &
         Spot VM+Harvested &
         Compute allocation
         \\
         Overclocking &
         CPU frequency &
         11\% less cost, $\uparrow$ Perf & %
         Regular VM &
         Regular VM+OC time &
         Reliability, power/energy
         \\
         Underclocking &
         CPU frequency &
         1\% less cost, $\downarrow$ Carbon &
         99\% Regular VM &
         Regular VM &
         Power, energy
         \\
         Non pre-provision &
         Spare compute &
         2\% less cost &
         98\% Regular VM &
         Regular VM &
         Compute allocation
         \\
         Region agnostic &
         Compute &
         22\% less cost, $\downarrow$ Carbon &
         \multicolumn{2}{c}{Region price} &
         Efficient region
         \\
         VM oversubscription &
         Compute &
         15\% less cost, $\downarrow$ Carbon &
         \multicolumn{2}{c}{85\% Regular VM} &
         Compute allocation
         \\
         VM rightsizing &
         Compute &
         50\% less cost, $\downarrow$ Carbon &
         \multicolumn{2}{c}{Rightsized VM} &
         Compute allocation
         \\
         MA datacenters &
         CPU frequency &
         40\% less cost &
         \multicolumn{2}{c}{60\% Regular VM} &
         Infrastructure cost \\
         \bottomrule
    \end{tabular}
    }
    \caption{Benefits and incentives of the cloud platform optimizations for the users.}
    \label{tab:optimization-benefits}
\end{table*}

\lexiang{Reviewers ignored this section and treat it as background info, thus they don't understand why hints are fundamentally helpful. We need to crisply describe the RQs at the beginning of this section and the intro to catch attention.}
\inigo{I also changed the title.}
\lexiang{Specified RQs in the intro paragraph of section 2.}

\review{Section 2 is nice but nearly all of it has little to do with the rest of the paper (which only considers 3 choices). It seems to be focused only on the choices provided by a singe public cloud. Reading through section 2, it is not clear what is relevant to the rest of the paper and what is just attempting to be comprehensive. Table 2 has lots of detail but again, not sure of the point and how much one needs to understand.}

\edit{We answer 2 research questions:
(1) What are the characteristics and requirements of cloud workloads?
(2) What workload characteristics are required to enable cloud optimizations?}

\subsection{Workloads}

\myparagraph{Methodology}
We study the characteristics of a diverse set of cloud workloads running at \CompanyX.
\edit{
We surveyed all the 990 internal workloads and got responses from 188 of them.
This represents 19\% of the workloads and 1.4 million cores across over 400K VMs.
}
These include web search, collaboration and productivity suites, and real-time communication workloads. They are deployed across 48 regions worldwide and used by hundreds of millions of users.

\myparagraph{Results}
We divide the results by workload characteristics, weighted by core usage.
\Cref{tab:survey-results} shows the core usage for each  characteristic. We group these into four main categories: scalability, reliability, performance, and geography sensitivity. 

First, under \emph{scalability}, we characterize workloads by their \emph{statelessness} (\ie, feasibility to scale in/out without persisting states) and \emph{deployment time requirements} (\ie, whether VMs have strict deployment latencies).
According to the survey, 62.9\% of the workloads are partially to fully stateless and the majority does not have strict deployment time requirements.

Second, under \emph{reliability}, we look at \emph{availability} (\ie, cloud downtime tolerance)
and \emph{preemptibility} (\ie, ability to pause and resume progress if X\% of the VMs are still alive).
The survey responses show that 62.8\% of the cloud workloads require three nines of availability or less, which translates to 8.76 hours to 36.5 days system downtime tolerance per year~\cite{googlesre}.
Besides, 60.6\% of the cloud workloads are at least partially preemptible.
These provides flexibility for cloud to manage resources more efficiently.
\edit{Workloads with 100\% preemptibility or 0\% availability requirement are generally test and dev environments.}

Third, \emph{delay tolerance} shows the workload flexibility within a specified deadline.
\edit{For example, a workload serving requests may have a tail latency service level objective (SLO) of 100 ms while most requests complete within 20 ms.
The specific target metric depends on the workload.
}
Our data indicates that around a quarter of the cloud workloads are tolerant to delays and they have a less strict performance requirement for the cloud platform.

Lastly, \emph{region independence} shows the workloads' ability to migrate among geographical regions without restrictions such as workload dependencies and security policies, and 61.4\% of the workloads are partly to fully available to migrate without negative impact on their operation.
\lexiang{We need to ensure consistency between ``region-independence"  and ``location".}
\inigo{I guess "region independence" is more correct.}

\subsection{Cloud optimization mechanisms}
\label{sec:systems}

Cloud platforms implement many optimizations to improve their efficiency.
To make our discussion concrete, we now look at ten common cloud optimizations \edit{among public cloud providers}. 
\Cref{tab:optimization-benefits} summarizes the resources, workload owner benefits \edit{(\ie, how much workload owners can save)}, pricing \edit{(based on public cloud pricing)}, and the platform benefit model \edit{(\ie, how does the platform benefit)} for each cloud platform optimization.
The rest of this section explains the mechanism, interface, and required workload characteristics for each optimization.

\myparagraph{Auto-scaling}
To allow workload owners to not always provision VMs for the peak load, providers offer auto-scaling to dynamically adjust the number of VMs based on load~\cite{mao2011auto}.
This allows owners to save money by running fewer VMs when not needed and providers to monetize this free capacity.

Auto-scaling is usually offered as a separate service~\cite{ec2-autoscaling,azure-autoscaling}.
Workload owners define their own policy defining a time schedule (\eg, scale out from 1 to 4 PM) or a load threshold (\eg, scale out if the CPU utilization is higher than 40\%) and it is suitable for workloads that allow scaling in and out, which is characterized by their stateless and delay tolerant nature.
Owners also need to specify \emph{deployment time requirement}, if the workload requires a VM to be immediately available.

\myparagraph{Spot VMs}
To monetize unallocated capacity, providers offer VMs with relaxed SLOs.
These VMs are evicted if their resources are needed by on-demand VMs.
Spot VMs are offered at discounted prices which allows owners saving money to run their workloads.
Providers usually offer Spot VMs~\cite{ec2-spot,azure-spot,google-spot} as a VM type or deployment flag and may offer dynamic pricing to decide which Spot VMs to evict first. %
Spot VMs are ideal for workloads that tolerate evictions~\cite{azure-batch-low,aws-batch-spot}.
These are workloads that support preemptions (\ie, 20\% or higher).

\myparagraph{Harvest VMs}
To use unallocated resources, cloud platforms can place more Spot VMs.
However, it's inefficient to create/remove VMs to use all the resources in a server.
Harvest VMs build on top of Spot VMs and can dynamically grow and shrink to utilize spare CPU~\cite{ambati2020slo,wang2021smartharvest}, memory~\cite{fuerst2022memory}, and storage~\cite{reidys2022blockflex} in the server.
This is similar to Burstable VMs~\cite{azure-burstable,ec2-burstable} but without the credit abstraction.
Providers offer harvesting as a new VM type or a deployment flag specifying the amount of resources to harvest~\cite{ambati2020slo}.
In addition to the characteristics from Spot VMs (\ie, high preemptibility), Harvest VMs are ideal for workloads that can scale up/down and thus they need to tolerate delays.

\myparagraph{Overclocking}
To improve workload performance, cloud platforms can increase component-level (\eg, CPU cores) frequency for VMs~\cite{jalili2021cost}.
To provide the benefit, the platform needs to determine the bottleneck resource to overclock and factor the impact of overclocking on component reliability and power draw in its decision-making.
The capability is provided to workload owners via dedicated VM types.
Workloads can also use an interface similar to auto-scaling to define a time schedule or a load threshold as signals to the platform for their overclocking requirements.
This optimization targets workloads with high CPU utilization periods~\cite{jalili2021cost} (\ie, $95^{th}$ percentile of max CPU utilization greater than 40\%), that can scale up/down, and are tolerant to delays.
\edit{
Overall, owners can provision fewer VMs to serve their peaks using overclocking.
}

\myparagraph{Underclocking}
Platforms can reduce their energy usage and carbon footprint by decreasing the frequency of VMs during periods of low activity.
This optimization is available through certain VM types offered by providers, similar to overclocking.
Workloads that are delay-tolerant, support scaling down, or have no persistent state
and can handle delays are well-suited for this optimization.

\begin{table*}[t!]
    \centering
    \scriptsize{
    \begin{tabular}{c|rcccccccc}
        
         \toprule
         \multirow{3}{*}{\makecell{\textbf{Cloud}\\\textbf{Optimization}}} & \multirow{3}{*}{\makecell{\textbf{Cores}\\(\%)}} & \multirow{3}{*}{\makecell{\textbf{Existing}\\ \textbf{Ad-hoc Interface}}} & \multicolumn{7}{c}{\textbf{Required Workload Characteristics}}\\
         \cline{4-10}
         &&& Scale      & Scale  & Deploy   &              &                 & Delay       & Region \\
         &&& up/down    & out/in & time     & Availability & Preemptibility  & tolerance   & independence \\
         \midrule
         Auto-scaling     &33.1&Dedicated service&            & \checkmark & \checkmark &              &                & \checkmark &             \\
         Spot VMs         &21.6&\makecell{VM type, deployment flags,\\dynamic pricing}&            &            &            &              & \checkmark     &            &           \\
         Harvest VMs      &6.4&VM type, deployment flags&   \checkmark&            &            &              & \checkmark     &      \checkmark     &              \\
         Overclocking        &41.3&VM type&  (\checkmark)     &            &            &              &     & \checkmark &              \\
         Underclocking       &36.0&VM type&  (\checkmark)     &            &            &              & \checkmark     & \checkmark   &            \\
         Non pre-provision    &68.8&Inferred&            &            & \checkmark &              &                &  &             \\
         Region-agnostic  &43.0&Explicit region selection&            &            &            &              &                &             & \checkmark \\
         VM oversub       &7.6& VM type, inferred&  (\checkmark)          &            &            &              &                & \checkmark  &            \\
         VM rightsizing     &2.1&Inferred, recommendation& (\checkmark) &  &         & \checkmark   & (\checkmark) &        &                 \\
         MA datacenters            &59.6&Inferred&            &            &            & \checkmark   &                &          &               \\
         \bottomrule
    \end{tabular}
    \caption{Overview of popular cloud optimizations, with percentages of applicable cores in the cloud platform, existing cloud interfaces, and required workload characteristics.
    To calculate the core percentages, we sum up the percentage of cores from workloads in the internal survey that has a specific workload characteristic.
    ($\checkmark$) indicates optional characteristic.
    \edit{
    }
    }
    \label{tab:cloudopt}
    }
\end{table*}

\myparagraph{VM pre-provisioning}
To reduce the time to create a VM, providers may provision VMs ahead of the time being instantiated when requested by workloads, hence, reducing the time to deployment~\cite{yao2021pps}.
This is a good complement to auto-scaling as it allows adding VMs quickly when needed (\eg, a load spike).
But cloud providers currently provision VMs without considering their utility to the workload they serve.
Disabling VM pre-provisioning (\ie, Non pre-provisioning) for workloads without strict deployment time requirements can reduce costs with minimal impact on performance.

\myparagraph{Region-agnostic placement}
Running workloads on VMs in cheaper and greener regions (\ie, regions with lower $CO_2$ emissions) can help reduce costs and carbon footprints, especially for workloads that do not have strict latency or data-locality requirements.
Currently, cloud providers require workload owners to specify the region for VM deployment.
Although there have been proposals for semi-automatic region selection~\cite{adnan2012energy,shi2022characterizing}, no commercial solutions are available yet.
Providers can place/migrate workloads to cheaper and greener regions (\eg utilizing solar energy) when needed if the  workloads are region-agnostic.

\myparagraph{VM oversubscription}
To increase server utilization, cloud platforms may oversubscribe servers by deploying more VMs on them than the available resources, relying on statistical multiplexing to manage resource allocation.
However, if all VMs spike at the same time, the platform will throttle the least critical VMs to ensure stability.
Currently, platforms heuristically determine which VMs can be oversubscribed and to what degree~\cite{cortez2017rc} or offer oversubscribed VM types explicitly~\cite{gcpe2}.
Further knowledge of the workload characteristics can help identify good candidates for oversubscription.
If the $95^{th}$ percentile CPU utilization of a workload is less than 65\% and the workload is delay-tolerant or non-user-facing, then it is suitable for oversubscription~\cite{cortez2017rc}.

\myparagraph{VM rightsizing}
To enhance efficiency and minimize expenses for workload owners, the platform provides smart VM selection
by identifying VM miss-utilization and recommend transitions to more suitable types/sizes.
Automated adjustments apply to preemptible workloads with relaxed availability requirements.
This is advantageous for workloads capable of scaling down less utilized components (\eg, below 50\%), facilitating a move to a smaller VM, typically half the original size. 
Conversely, if a single resource encounters high usage, the VM type can be upgraded. 
Overall, optimal VM selection considers factors like workload resource needs, performance requirements, and budget constraints.

\myparagraph{Multi-availability datacenters (MA DCs)}
Cloud providers can reduce infrastructure redundancy (\eg, power delivery and cooling) to decrease costs.
However, this may lead to infrastructure failures or maintenance events that require the platform to throttle or selectively turn off servers.
Traditionally, platforms have inferred which VMs are less critical and throttled them down or evicted them.
MA DCs take further advantage of workloads that explicitly require low availability, providing resources and charging users accordingly.

\section{Revisiting the cloud abstraction}

From the descriptions of the optimizations above, we observe that each one relies on a different subsets of workload characteristics, and has particular interfaces to gather %
inputs from workload owners.
\Cref{tab:cloudopt} summarizes the required workload characteristics and interfaces for each optimization.

Problems with these interfaces are four-fold:
(1) They are ad-hoc: some are specific to a service, some rely on VM type and deployment flags, some rely on inference, or are just based on recommendations.
As cloud platforms introduce new optimization mechanisms, the overall cloud interface becomes complex and untenable.
For example, \emph{Spot VMs}, \emph{Harvest VMs}, \emph{Overclocking}, and \emph{Underclocking} each require an additional dimension in the already-complicated VM type interface, which limits their usage.
(2) These interfaces are tied to the corresponding optimizations and %
require expertise with the corresponding optimization. 
For instance, workload owners need to know Harvest VMs can shrink/expand core counts and are applicable if their workloads can scale up/down.
(3) They are mostly static, as they tend to be specified at deployment time.
(4), Many of these interfaces rely on inferred characteristics with questionable accuracy and lack of explicit user contracts.
As a result, the provider in many cases has to be conservative, and not utilize optimizations to their fullest extent.
For example, in the absence of extra information, the provider has to assume that a VM requires maximum reliability, and cannot move to other regions.

To address these challenges, we observe that ideally the cloud interface should \emph{decouple} workload characteristics -- which are
known to workload owners -- from the cloud optimizations they enable -- best understood by the cloud provider. This creates a proper separation of concerns, can reduce interface complexity and changes as optimizations evolve, and can enable the effective utilization of
cloud optimizations. In this paper we propose an extension to the cloud interface that enables this separation. Before describing our proposal in \Cref{sec:framework}, we first discuss some challenges
and requirements that such an extension should meet.

\subsection{Challenges}
\label{sec:challenges}
\myparagraph{Generality}
We need to support a wide range of workloads and cloud platform optimizations.
The interface needs to be \emph{general} for any workload to express their main characteristics and requirements.
These characteristics can also be dynamic and change over time.
In addition to the interface at deployment time, we need to expose an interface to allow updating the characteristics and requirements at runtime.

Based on the survey results and our internal discussions with cloud optimization teams, \Cref{tab:cloudopt} summarizes the essential workload characteristics that cloud optimizations need to operate.
Note that many of these characteristics benefit multiple cloud optimizations.
We target these ten cloud platform optimizations but this number may grow.
In addition, some workloads may introduce new characteristics.%
The interface needs to be \emph{extensible}.%

Critical optimizations (\eg, \emph{MA DCs}) require to push updates to the platform in real-time, while other optimizations (\eg, \emph{Spot VMs}) may pull information only when needed  (\eg, to create room for on-demand VMs).
The same applies to workloads.
For example, for \emph{Spot VMs}, we need to push priority updates and receive future eviction events.
Therefore, we need to provide both pull and push interfaces.

\myparagraph{Incentives} 
Workload owners must be incentivized to use any interface extensions.
It is clear from \Cref{tab:optimization-benefits} that the optimizations have the potential to reduce cost and/or improve performance, sustainability for workloads.
The interface implementation should guarantee that VMs are \emph{no worse off} by using the interface, and possibly better. 
Any extension should also be \emph{incrementally adoptable}, meaning that performance/cost/sustainability should not degrade by not adopting the interface, after its introduction. 

\myparagraph{Safety}
As public cloud platforms, safety must be ensured.
We need to prevent workloads from providing wrong information (\eg, due to bugs) and ensure its public interfaces are resilient to attacks (\eg, denial of service).
The interface should not enable workloads to abuse the system.%

For workload owners, 
we must protect against leaks of sensitive information from workloads by using interface isolation and encryption to prevent side-channel attacks.
At the same time, we need to ensure the \emph{correctness} of the information to prevent %
performance degradation or unnecessary cost.

\review{The authors discuss the importance of safety in public cloud platforms. There's a need to prevent users from providing incorrect information, which could be detrimental to the system. A detailed security analysis and discussion on how the safety mechanisms would be implemented in real-world scenarios will further improve this work.}

\myparagraph{Coordination}
The goal of our extension
is to enable the cloud platform to reason about the information it receives and optimize accordingly, while maintaining quality of service.
To achieve this, any implementation must aggregate data at different levels for various optimizations.
For example,
\emph{Auto-scaling} considers all the VMs for a workload,
\emph{Overclocking} considers physical domains (\ie, servers and racks),
and
\emph{Spot VMs} considers all the VMs that can be evicted.

We must also enable \emph{coordination} between multiple cloud platform optimizations that may want to take actions on the same resources.
This coordination is necessary to resolve conflicts that may arise when, for example, both \emph{Overclocking} and \emph{MA DCs} are attempting to adjust CPU frequencies simultaneously.
At the same time, we need to ensure the resources are shared \emph{fairly} among multiple workloads.

\subsection{Requirements}
\label{sec:requirements}

\inigo{Is there a better way to introduce this?}

Besides addressing the above challenges, any implementation of
a cloud interface extension must address:%

\myparagraphemph{Scalability} %
It must be \emph{scalable} and handle a high rate of dynamic bi-directional communication between many workloads and multiple cloud optimizations.
Potentially, it needs to support exchanging information from/to all VMs in the cloud platform.

\myparagraphemph{Availability} %
The use of any new interface must maintain \emph{high availability} and \emph{tolerate failures}. The new information provided must be 
\emph{persisted} even if cloud optimizations or workloads are restarted.

\myparagraphemph{Efficiency} %
\edit{The extension needs to be \emph{low-overhead} and avoid imposing unnecessary burdens and overheads on the system.}

\myparagraphemph{Maintainability} %
Maintaining a new service requires effort for development, operation, bug fixing, and others.
We need to build a \emph{simple} service with minimal maintenance overhead and rely on existing infrastructure as much as possible.

\section{Workload Intelligence}
\label{sec:framework}

Considering these challenges and requirements, we propose Workload Intelligence (WI) as an extension to the cloud interface: a framework for dynamic bi-directional communication between cloud workloads and the cloud.
WI allows workloads to explicitly specify their characteristics and requirements through \emph{hints} and to dynamically change them.
WI makes hints available to the cloud platform to optimize its operation.

\myparagraph{Hints}
They are \emph{best-effort}: there is no guarantee that they will be fulfilled, but they may (only) improve some dimension of quality: price, performance, sustainability, etc.
They are also \emph{incentive-compatible}: in their absence, the provider assumes the most conservative version of a workload characteristic.

\begin{figure}[t]
\centering
\includegraphics[width=\columnwidth]{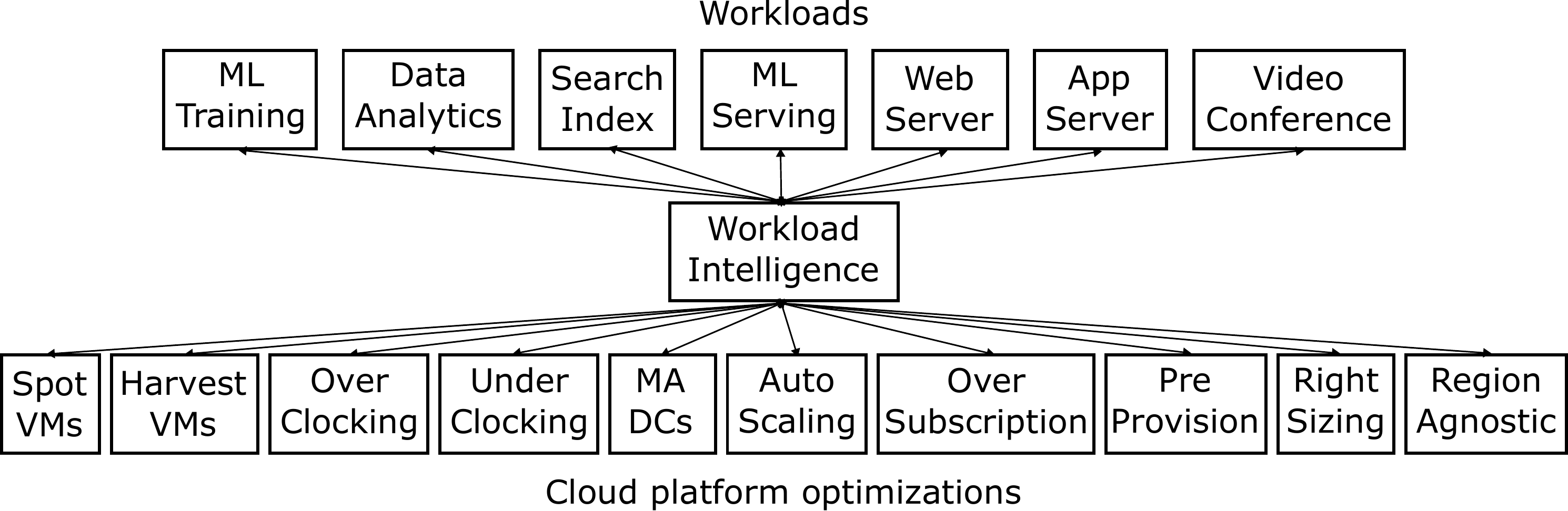}
\caption{Workload Intelligence overview.}
\label{fig:overview}
\end{figure}

With WI in place, the cloud platform can significantly simplify its offerings, reduce costs without fear of violating workload requirements, and lower prices for workload owners.
\Cref{fig:overview} shows an overview of WI enabling communication between multiple workloads and optimizations.

\inigo{We need to make the point that we are targeting true positives, and we don't want false positives where we enable something that we shouldnt have: https://shiffdag.medium.com/what-is-accuracy-precision-and-recall-and-why-are-they-important-ebfcb5a10df2}

\myparagraph{Workload hints}
\edit{
We define seven hints based on the workload characteristics needed by cloud optimizations identified in \Cref{tab:cloudopt}:
(1) scale up/down (boolean), (2) scale out/in (boolean), (3) deploy time (milliseconds), (4) availability (number of 9s), (5) preemptibility (percentage), (6) delay tolerance (milliseconds), and (7) region independence (boolean).}

\edit{
Hints define if that particular characteristic is relaxed (\eg, the workload has low availability requirements).
If unspecified, we assume the most conservative setting (\eg, the workload wants fast deployment times for its VMs).
Common workload targets and goals (\eg, cost and $CO_2$) are not hints.
}

\review{different workloads or different optimizations might define different tags for the same information, e.g., is availability expressed as a tag followed by number of nines, seconds of down time per year, etc. I would encourage the authors to describe how they address this and to include examples of the API used to provide hints.}
\lexiang{Added types for each characteristics to address the review.}

\myparagraph{Platform hints}
In the other direction, WI allows the platform to programmatically inform workloads about upcoming events and opportunities for optimization, among other useful scenarios.
\edit{Example hints include VM evictions for Spot VMs and higher CPU frequency available for overclocking.}
\edit{Workloads can then react to these hints by specifying a VM with the lowest penalty upon eviction for graceful shutdown or a VM with the highest benefit for overclocking.}

\lexiang{Need to also describe the WI API in whole.}
\inigo{We also need to talk about the ``language'' and how we make sure that all optimizations use a common interface. Including some examples.}
\review{
I am left wondering on how hints are concretely expressed such that they are extensible and interoperable between different optimization, i.e., what is the language for hints. There were also multiple places where I felt there was missing information to enable me to completely understand WI.
}
\inigo{I added the previous two paragraphs to clarify this.}
\lexiang{Added workloads reacting to platform hints in the second paragraph to complete the three types of hints.}

\subsection{Architecture}
\begin{figure}[t]
    \centering
    \includegraphics[width=\columnwidth]{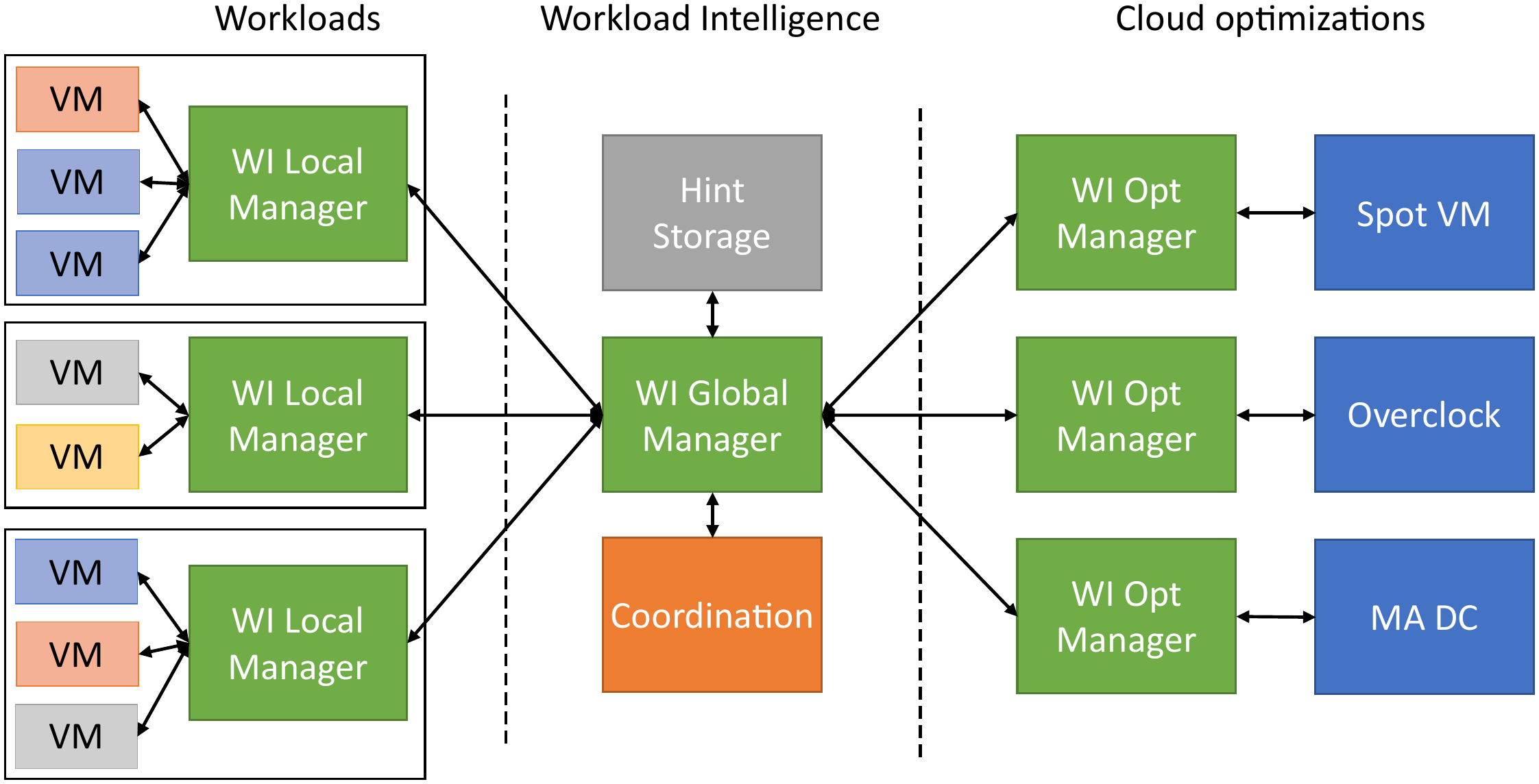}
    \caption{Workload Intelligence architecture showing three example cloud optimizations.
    }
    \label{fig:architecture}
\end{figure}

\Cref{fig:architecture} shows the WI architecture with an example for three cloud optimizations:
Spot VMs, Overclocking, and MA DCs.

\myparagraphemph{Local managers}
For \emph{scalability}, each server in the cloud provider runs a local WI manager as shown in the left of \Cref{fig:architecture}.
This local manager collects the runtime hints from the workloads running in the VMs in the server and passes them to the global manager.
This local manager also collects the notifications coming from the cloud optimizations and exposes them to the VMs running the workloads.

\myparagraphemph{Global manager}
For every region, we have a global WI manager that is logically centralized but physically distributed.
This is shown in the in the center of \Cref{fig:architecture}.
This component stores the hints, aggregates them, and enables \emph{coordination} across multiple cloud optimizations and multiple workloads.
It acts as a broker that exchanges information and hints between the cloud optimization and the workloads running in VMs.
It provides multiple interfaces to retrieve this information at scale in near real-time and aggregate it at multiple granularities (\eg, per server and per rack).

\myparagraphemph{Cloud optimization managers}
Each optimization can leverage a basic WI optimization manager.
This manager gets the hints from the global manager.
It also uses the global manager to pass hints to the workloads through the local manager.
The right of \Cref{fig:architecture} shows the example for three optimizations.

\subsection{Communicating and storing hints}

To provide \emph{scalability}, \emph{high availability}, 
\emph{maintainability}, %
and \emph{fault tolerance}, WI uses a combination of a PubSub and a distributed database to communicate and store the hints.
For the PubSub, WI uses Kafka~\cite{wang2015building} which synchronously delivers the hints at \emph{large scale}.
For the database, it uses \CosmosDB%
\footnote{Fictitious name for a cloud database, for anonymity.}
which provides \emph{fault tolerance} and 
\emph{durability}. These are also mature services that are \emph{easy to maintain}.

Depending on the use case, hints need to be sent synchronously or asynchronously.
For example, in the Spot VM case, workloads can specify their evictions preference asynchronously and the cloud platform gathers this information whenever it needs the capacity for regular VMs.
On the other hand, when the cloud platform decides to evict VMs, it needs to immediately notify the VM.

\myparagraph{Deployment hints}
When deploying a VM (or a set of VMs), the workload owner can specify the attributes for the workload that they will run (\eg, tag a set of VMs as highly preemptible).
Workload owners can specify these hints through the common deployment interfaces~\cite{aws-cloudformation,azure-arm,gcp-deployment}.
The cloud platform internally uses the WI global manager REST interface to store these hints.
The global manager stores the deployments hints in \CosmosDB and publishes them using Kafka.
Cloud platform optimizations can leverage the optimization manager to
(1) retrieve this hints asynchronously when needed or (2) subscribe to the hints of a particular type.
These two actions offer a \emph{general} interface that can be used by a wide variety of workloads.
Workloads can also update their hints after deployment while the VMs are running.

\begin{table}[t]
    \centering
    \scriptsize
    \begin{tabular}{cl}
    \toprule
    Priority & Cloud optimization \\
    \midrule
    0 & On-demand \\
    1 & MA datacenters \\
    2 & Rightsizing \\
    3 & Oversubscription \\
    4 & Auto-scaling \\
    5 & Non pre-provision \\
    6 & Region agnostic \\
    7 & Underclocking \\
    8 & Overclocking  \\
    9 & Spot VMs \\
    10 & Harvest VMs \\
    \bottomrule
    \end{tabular}
    \caption{Priorities across our ten cloud optimizations.}
    \label{tab:opt-priorities}
\end{table}

\myparagraph{Runtime hints: workload $\rightarrow$ platform}
The workloads running in the VMs can also provide hints.
For example, a VM that it is currently not running critical jobs can specify it is more tolerant to evictions.
These runtime hints can be specified within the VM or from a logically centralized manager.

For a workload running inside of a VM to set a runtime hint, it uses local interfaces (\eg, Hyper-V KVP~\cite{hyperv-kvp} or XenStore~\cite{xenstore}).
The local WI manager in each server polls for these runtime hints and uses Kafka to publish them.
The global manager is subscribed to these events and stores them in \CosmosDB.
The optimization manager can also subscribe to these events or asynchronously check through \CosmosDB.

In addition to the local hints that VMs can specify, a global workload manager can use the global manager REST interface to specify hints.
For example, the Resource Manager in a Hadoop YARN~\cite{YARN} deployment can set preference for evictions for a set of VMs.

Multiple entities can be publishing hints for the same resource.
To provide \emph{correctness}, if the cloud platform optimization identifies that the hints are not consistent, it can notify the workload that it is ignoring them.

\myparagraph{Runtime hints: platform $\rightarrow$ workload}
To let workloads adapt to the cloud platform actions, such as evictions of Spot VMs, WI provides a mechanism to send hints (\eg early notifications) from the cloud optimizations to the workload via Kafka.
The global manager is subscribed to these events and stores these hints in \CosmosDB.
The local manager also subscribes to these events and exposes them to the VM through the local interfaces.
Cloud platforms already offer interfaces for VMs to locally check their attributes.
An example is the metadata service~\cite{gcpmetadata,azuremetadata,awsmetadata} which offers data like the VM identifier or the type of VM.
Scheduled events~\cite{azurescheduledevents,awsscheduledevents} is another communication channel which notifies about events that will happen soon (\eg, reboot, maintenance, evictions).

The cloud platform can also inform the workload that a planned maintenance event will take place.
With this info, the workload can shut down gracefully.

\subsection{Providing hints safely}
\edit{WI ensures safety for both workload owners and the cloud platform.}
To protect the interface against DoS attacks, we enforce maximum rates per optimization and workload when setting deployment and runtime hints for all interfaces separately.
As hints are best-effort, DoS mitigations are simpler.

To protect the cloud platform from abusive usage of a single resource, the cloud enforces fair-share among VMs and between workload owners.
Also, the cloud platform has the right to provide alternative optimizations (\eg, Spot VM vs. VM Pre-provisioning) to suit workloads' need.

To protect workload owners from side-channel attacks, we encrypt the hint communication.
Besides, the cloud platform does not provide details on its resource management decisions after applying (or ignoring) hints to prevent information leaks (\eg, VM placement).
For example, in the case of a Harvest VM expansion, the cloud platform does not give reasons on their decisions and only the target VM is directly informed.

To protect the owners from sending wrong information, the cloud platform ignores any inconsistent/incompatible hints based on history.
In addition, from the incentives point of view, workloads can only hurt their own performance and cost by providing wrong information.

\review{The authors discuss the importance of safety in public cloud platforms. There's a need to prevent users from providing incorrect information, which could be detrimental to the system. A detailed security analysis and discussion on how the safety mechanisms would be implemented in real-world scenarios will further improve this work.}
\inigo{Doesn't the previous text cover this? The real-world part would be missing I guess.}
\lexiang{This reviewer wants implementation details of the security mechanisms, which is not the main focus of this paper.}

\subsection{Coordinating optimizations}
\label{sec:coordination}

\begin{figure}[t]
    \centering
    \includegraphics[width=0.45\columnwidth]{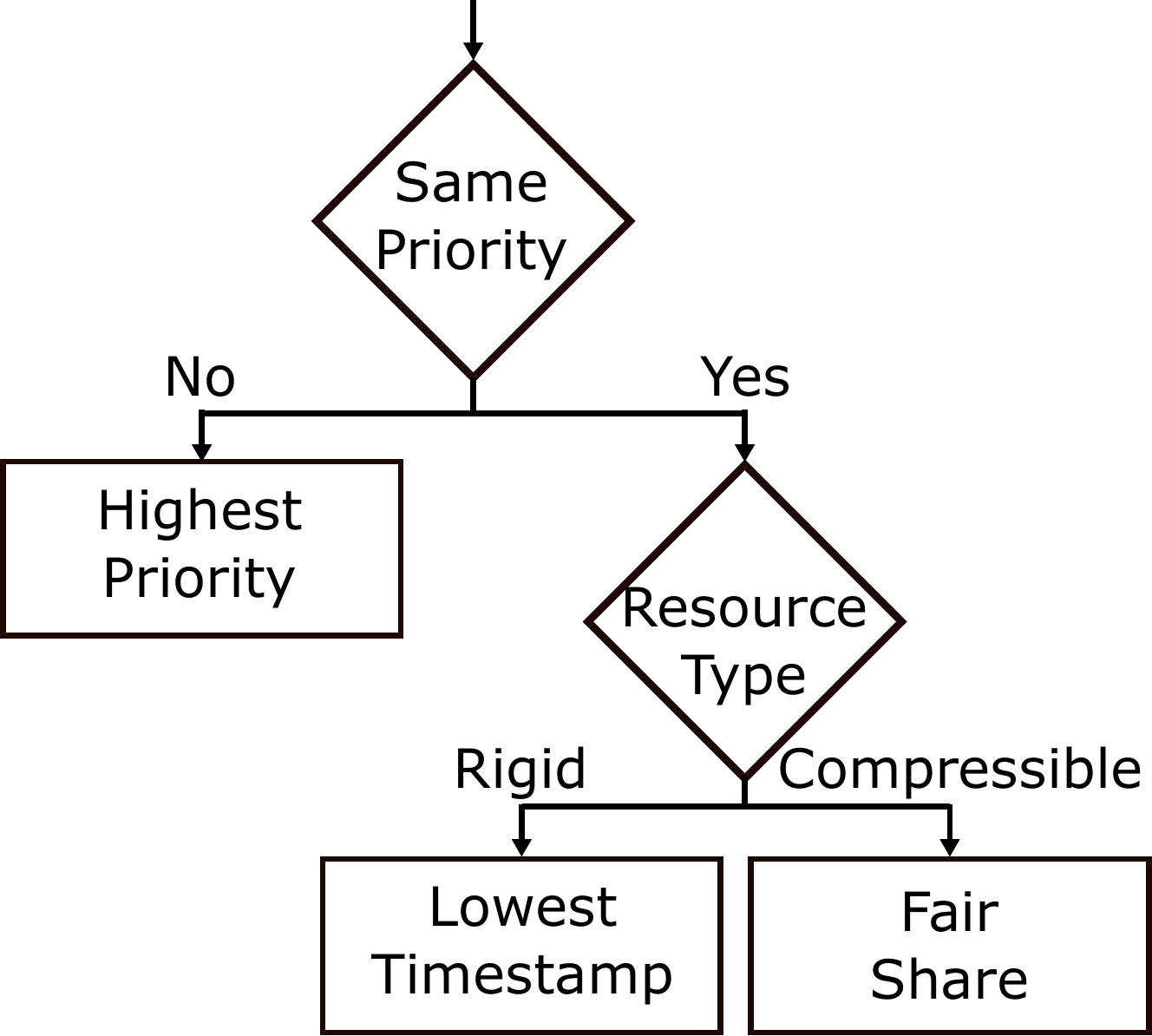}
    \caption{Conflict resolution for competing resources.}
    \label{fig:conflict-resolution}
\end{figure}

WI enables the cloud platform to reason about the information it receives to maximize its opportunities for optimization while protecting quality of service.
This includes handling conflicts that may arise when multiple cloud optimizations target the same resources.
For example, both Spot VMs and Pre-provisioning try to use the unallocated capacity in servers.
Another example, \emph{Overclocking} may try to increase the CPU frequency~\cite{jalili2021cost} and \emph{MA} may try to reduce it~\cite{jalili2021cost,zhang2021flex}.

To address potential conflicts, we implement an algorithm based on cloud optimization priorities.
\Cref{tab:opt-priorities} shows our priorities for the ten discussed optimizations (on-demand VMs have the highest priority).
This ranking reflects their significance for the provider, with Harvest VMs being the most opportunistic.
Note, most optimizations do not compete for the same resources \edit{and many of them (\eg VM rightsizing) are designed to release unnecessary resource} (\Cref{tab:optimization-benefits}).

\Cref{fig:conflict-resolution} illustrates this algorithm.
In cases where the optimizations have the same priority but the resource is compressible (\eg, CPU), we employ \emph{fair} share allocation.
Otherwise, we allocate the resource to the VM with the earliest request time. 
In the rare event of requests submitted simultaneously, we pick randomly.
Moreover, the WI ensures the fair sharing of resources among multiple VMs, guaranteeing equitable distribution between workloads.

\subsection{Alternative designs}
We discuss trade-offs between several design choices with a focus on coordination policy and interface.

\myparagraph{Coordination}
To resolve conflicts when the cloud platform attempts to apply multiple optimizations on the same resources, one can envision three approaches.

\myparagraphemph{Pricing}
The cloud platform specifies a price for each of the optimizations.
It can then use this pricing to resolve conflicts.
However, each optimization has different magnitudes, making it challenging to unify prices into the same unit.
Additionally, the price can change over time and may not always reflect the intrinsic value of the resources. (\eg, special discounts).

\myparagraphemph{Bidding}
Workload owners can define a price they are willing to pay for their VMs.
The cloud platform can then use this information to resolve conflicts and assign resources to the highest paying workload owners.
This is similar to bidding pricing strategy for Spot VMs, which has been either dropped or unsupported by major cloud providers~\cite{spotbidding, google-spot, azure-spot}.
Fundamentally, this approach is challenging for workload owners to understand and could introduce a gaming aspect.

\myparagraphemph{Our approach}
When building a new cloud optimization, we define a priority for each optimization and apply a set of rules depending on the resource in conflict.
This approach allows for easy unification and is simple for workload owners to understand and reason about.
At the same time, it makes it easy for the platform to operate and maintain.

\myparagraph{Interface}
To offer cloud optimizations to workload owners, one can envision three major interface designs.

\myparagraphemph{Reduced}
To reduce the complexity of the cloud systems and increase the generality of the interface, one can simplify the interface to take no workload owner inputs.  
For example, instead of letting owners to specify the VM types and required resources, the platform can infer the workload characteristics from the owners' past history and provide a VM type based on its best guesses. 
This approach works well with predictable workloads (\eg, scheduled events), however, it does not support optimizations to react to workload changes in time (\eg, delayed auto-scaling, sub-optimal Spot VM evictions).

\myparagraphemph{Discrete}
The current design maintains discrete interfaces for each individual optimization. 
This preserves the most flexibility for individual systems, which could potentially lead to well-optimized cloud operations. 
However, the predominant drawback is its sheer complexity, since each interface is customized. 
Besides, this design is not extensible, as the number of interfaces explodes as new cloud optimizations emerge. 

\myparagraphemph{Centralized}
A key observation from the discrete design is that many of the existing interfaces require similar workload characteristics from the workload owners either directly or implicitly. 
Instead of letting owners indirectly or repetitively submit their requirements, we can create a centralized interface that aggregates the minimum set of workload information needed and store them at one place. 
This design has lower complexity, but the communication overhead is potentially high, since all workloads and optimizations need to communicate via the centralized storage.

\myparagraphemph{Our approach}
To combine the benefits of both the discrete and the centralized design, we propose a hybrid interface.
In our design, the hints are physically distributed among workloads and optimizations (\ie, WI Local Managers and WI Opt Managers), to provide flexibility to update their runtime status at desired frequency. The WI Global Manager aggregates and distribute hints at one place (\ie, logically centralized). This allows optimizations and workloads to exchange hints on demand.
In addition, the workload characteristics utilized by WI are shared among multiple cloud optimizations and multiple workloads. To onboard future workloads/optimizations, the WI interface only needs to be updated for the delta, if any.

\subsection{Other resources}
\edit{
While current cloud optimizations primarily focus on compute resources, WI has the potential to benefit other resources (\eg, storage and networking).
For storage, \emph{delay tolerance} hints enable opportunities for co-locating storage and compute for workloads with I/O bottlenecks or using cheaper storage for lower costs in delay-tolerant workloads.
\emph{Region independence} hints can indicate data locality requirements and security concerns (\eg, GDPR), which can help enforce the desired data replication configurations.
}

\edit{
For networking, cloud Load Balancer services can benefit from \emph{scalability} and \emph{availability} hints to make better task placement decisions.
}
\edit{Also, \emph{delay tolerance} hints can be used in future optimizations to adjust cloud Content Delivery Network (CDN) service levels to reduce costs for delay-tolerant workloads or improve performance for delay-sensitive workloads.
It can be used in conjunction with \emph{region-independence} hints to optimize which regions to cache the data.}

\section{Implementation}

\subsection{Extending workloads beyond VMs}
\label{sec:impl-workloads}
Many workloads leverage frameworks and orchestrators like Kubernetes~\cite{kubernetes} and Functions-as-a-Service (FaaS)~\cite{shahrad2020serverless} for their deployment.
Cloud providers offer these managed frameworks~\cite{aks,eks,azurefunctions,awslambda} and they are usually deployed on top of VMs~\cite{aks,eks,shahrad2020serverless,agache2020firecracker}.

These frameworks can leverage WI to orchestrate the workloads running on top or directly expose WI. %
This enables workloads running on these frameworks to take advantage of cloud optimizations with minor extensions.
In this section, we describe how to extend three common frameworks.

\review{Furthermore, the design assumes a virtual machine is the rental unit for users. It would be good to discuss other types of clouds that provide containers and functions for end users.}
\inigo{I changed the title of the section but this should be clear enough.}

\myparagraph{Big data analytics: Hadoop}
To support WI, we extend Harvest Hadoop~\cite{ambati2020slo,fuerst2022memory}.
The management components (\eg, Resource Manager and Name Node) run on VMs with high requirements while the workers run in a mix of VMs with low and high requirements. %

We use the WI interface to retrieve notifications for evictions and change of resources (\eg, more CPU or memory available).
Then we pass this information to Harvest Hadoop which already handles evictions and changes in the number of resources (\eg, CPU and memory).

We also add a new component to the workers that uses the local WI interface to specify the VM priority depending on:
the criticality, the amount of containers running, elapsed job processing time, and whether containing master nodes.
For example, a VM that is running many critical containers will have a ``High'' priority while an empty node will have ``Low''.
This will make ``High'' priority VMs less likely to be evicted and to get more resources (\eg, harvesting or overclocking).

\myparagraph{Microservices: Kubernetes}
The Kubernetes control plane components run in VMs with high requirements (\eg, high-availability, low preemptibility).
When provisioning the worker nodes, we leverage Karpenter~\cite{karpenter}, a node provisioning manager for Kubernetes clusters.
We extend Karpenter to provide hints based on the pod requested by the applications.
The worker nodes can be a mix of Regular, Spot, Harvest, Overclocking, Underclocking, and Oversubscribed VMs.
These workers are grouped into different pools based on their characteristics.
We also add a new component to each VM which runs next to the Kubelet and uses the WI interface to provide and receive hints.
For example, if a VM is to be evicted, the WI component uses graceful shutdown to stop the pods in that VM and migrate the load to other pods.

Workloads running on Kubernetes use the node pool abstraction through tolerations and node affinities~\cite{k8s-scheduling}.
For example, if we want to run the Social Network from the DeathStarBench~\cite{gan2019open} on Kubernetes~\cite{qiu2020firm}, we specify the frontend pods to run in the node pool with high-preemptibility.
In addition, the logic microservices (\eg, compose post, social graph, write timeline) can specify when their latency is too high and trigger optimizations like overclocking.

\myparagraph{FaaS: OpenWhisk}
We use the OpenWhisk implementation from FaaS on Harvest VMs~\cite{zhang2021faster} as a base.
Similarly, to Hadoop and Kubernetes, we run the logically centralized control plane components (\eg, Nginx, Controller, Kafka) on VMs with high requirements (\eg, low fault tolerance and high-availability).
For the worker components, we use VMs with heterogeneous requirements based on the workload.
We add a component to track the running functions and adjust the characteristics of the VMs that are running worker nodes.
For example, if we have many long running functions, we deploy more regular VMs while if the functions are mostly short, we deploy them on VMs with lower requirements.

In the worker node, we extend the \emph{Resource Monitor} that runs next to the \emph{Invoker} to interact with the local WI interface.
Depending on the number of functions running on the worker and their duration, the Resource Monitor sets a higher or lower priority for the VM.

\myparagraph{Other workloads}
The extensions for these three frameworks can be implemented for workloads that do not leverage orchestrators.
Even without the runtime extensions, some workloads can leverage deployment hints by tuning the way they are deployed.
For example, the workload owner could specify that specific VMs can leverage scaling up and WI can apply corresponding optimizations such as overclocking.

\begin{table}
    \centering
    \scriptsize
    \begin{tabular}{cl}
    \toprule
    \textbf{Optimization} & \textbf{Modifications for WI} \\
    \hline
    Auto-scaling     & Consume deployment \emph{scale in/out} hints. \\
    \hline
    Spot VMs         & Consume deployment \emph{preemptible} hints. \\
                     & Consume runtime \emph{preemption} priority. \\
                     & Publish runtime \emph{preemption} notification. \\
    \hline
    Harvest VMs      & Same as Spot VMs. \\
                     & Consume runtime \emph{scale up/down} priority. \\
                     & Publish runtime \emph{scale up/down} notification. \\
    \hline
    Overclocking     & Consume deployment \emph{scale up/down} hints. \\
                     & Consume runtime \emph{scale up} priority. \\
                     & Publish runtime \emph{scale up} notification. \\
    \hline
    Underclocking    & Consume deployment \emph{scale up/down} hints. \\
                     & Consume runtime \emph{scale down} priority. \\
                     & Publish runtime \emph{scale down} notification. \\
    \hline
    Pre-provision    & Consume deployment \emph{deployment time} hints. \\
    \hline
    RA placement     & Consume deployment \emph{locality} hints. \\
    \hline
    Oversubscription & Consume deployment \emph{scale up/down} hints. \\
                     & Consume deployment \emph{delay tolerance} hints. \\
                     & Consume runtime \emph{scale down} priority. \\
    \hline
    VM rightsizing   & Consume deployment \emph{scale up/down} hints. \\
                     & Consume deployment \emph{delay tolerance} hints. \\
    \hline
    MA DCs           & Consume deployment \emph{scale up/down} hints. \\
                     & Consume deployment \emph{preemptible} hints. \\
                     & Publish runtime \emph{scale down} notification. \\
                     & Publish runtime \emph{preemption} notification. \\
    \bottomrule
    \end{tabular}
    \caption{Extensions to existing cloud platform optimization.}
    \label{tab:opt-wi-impl}
\end{table}

\begin{table*}[t!]
    \centering
    \scriptsize{
    \begin{tabular}{c ccccccc}         
        \toprule
         \multirow{3}{*}{\makecell{\textbf{Case study}}} & \multicolumn{7}{c}{\textbf{Workload characteristics of required hints}}\\
         & Scale up/down     & Scale out/in     & Deploy time    & Availability  & Preemptibility & Delay tolerance  & Region independence \\
         \midrule
        Big data analytics (\S\ref{sec:bigdata})    & \checkmark  & \checkmark & &  & \checkmark &  \checkmark & \\
         Microservices (management nodes) (\S\ref{sec:microservice}) & \checkmark & & & \checkmark &  &   & \\
         Microservices (worker nodes) (\S\ref{sec:microservice}) & \checkmark & \checkmark & \checkmark & \checkmark & \checkmark & \checkmark  & \\
         Video conference (media-service VMs) (\S\ref{sec:videoconf}) & (\checkmark) & \checkmark & \checkmark & \checkmark & \checkmark & \checkmark &  \checkmark \\
         \bottomrule
    \end{tabular}
    \caption{Overview of workload specified hints utilized by WI for our case studies.
    ``$\checkmark$" indicates hint required.
    }
    \label{tab:casestudy}
    }
\end{table*}

\subsection{Extending cloud platform optimizations}

For cloud platform to onboard an optimization, we need to define (1) managed resources and the (2) priority compared to other optimizations.
In addition, to track the benefits we also need to define the (3) workload owners benefit, (4) pricing, and (5) cost model.
This is not difficult for new optimizations to get onboard, because all these features should have been defined by the optimizations themselves.

For existing cloud optimizations,
\Cref{tab:optimization-benefits} defines the resources and the pricing for each optimization,
\Cref{tab:opt-priorities} specifies the priorities for each optimization, 
and \Cref{tab:opt-wi-impl} describes the hints that they need to consume and publish from and into WI.
When applying the changes to the resources, the cloud optimizations leverage the coordination described in \Cref{sec:coordination} which leverages priorities.

\review{The system has been evaluated using a few use cases but has not been deployed practically in the cloud. It would be good to delve deep into the practical challenges of deploying such a system.}

\section{Evaluation}
To evaluate WI, we first describe three use cases that demonstrate how three workloads can leverage multiple cloud platform optimizations.
We choose these three workloads because they fall into the categories of big data analytics, web applications and real-time communication respectively, and together, these workload classes comprise 84\% of the cloud cores usage at \CompanyX.
In addition, \Cref{tab:casestudy} shows that the workloads for our evaluation provide both good coverage and high diversity of hints.
We then evaluate the potential for cost and carbon savings at a cloud provider scale.

\subsection{Case study: Big data analytics}
\label{sec:bigdata}
\myparagraph{Methodology}
We deploy our WI-aware Hadoop (\S\ref{sec:impl-workloads}) on a 20-node cluster composed of 5 VMs for the management components (\ie, Resource Manager and NameNode) each with 4 vCPUs and 16 GB of memory and 15 VMs for the workers each with 8 vCPUs and 64 GB of memory.
\Cref{tab:casestudy} summarizes the hints for deploying the worker node VMs.

We use a 5-day MapReduce workload trace from a production cluster at \CompanyX in June 2020.
For reproducibility, we scale the trace down to fit into a 20-node cluster by randomly down-sampling at a 2\% rate.
This scaled down trace comprises 100 jobs and lasts 5 hours.
To emulate the characteristics of the original jobs, we assign the same job priorities to our synthetic MapReduce jobs as in the original trace.

\edit{For reproducibility, we will open-source the setup to emulate this experiment in \CompanyX. This includes a ``user-space'' implementation of WI.}

\myparagraph{Operation}
Using the WI deployment hints, 
the platform decides to enable Auto-Scaling, Spot VMs, and Harvest VMs for the VMs running the Hadoop workers.
\edit{
At runtime, each Hadoop worker posts hints to the local WI server with the runtime ``preemptibility`` hint for that VM every second \edit{(same as the default Apache YARN heartbeat interval~\cite{YARN})}.
To calculate this hint, the workload consider the number of containers, their uptimes, user-specified job priorities, and whether it hosts Application Masters.
For example, if the worker VM has been running many critical jobs for a long time (\eg, >30 seconds based on the typical cloud eviction notice time), it unmarks the runtime ``preemptibility`` hint to reduce the eviction probability and maximize the amount of resources assigned.
WI uses these hints to determine which VMs to shrink and evict when reclaiming resources.
}

\begin{figure}[t]
\centering
\includegraphics[width=0.99\columnwidth]{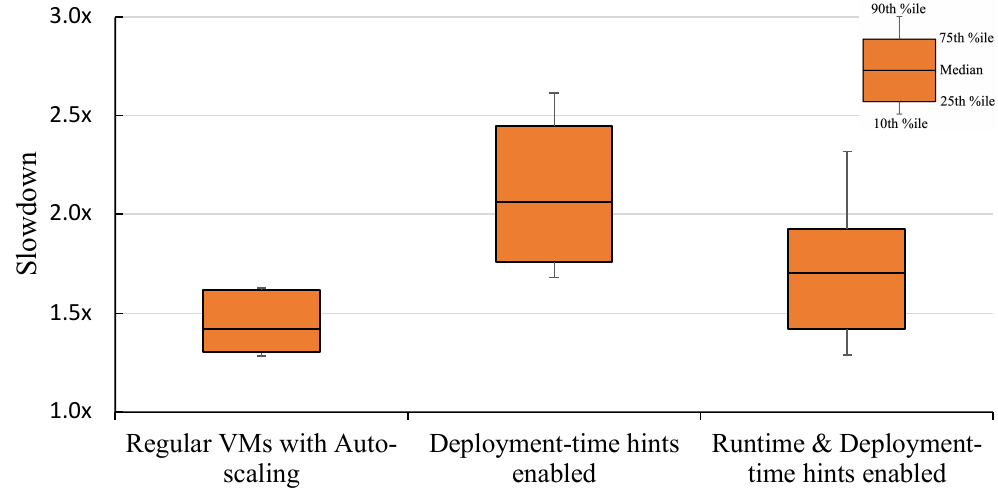}
\includegraphics[width=0.99\columnwidth]{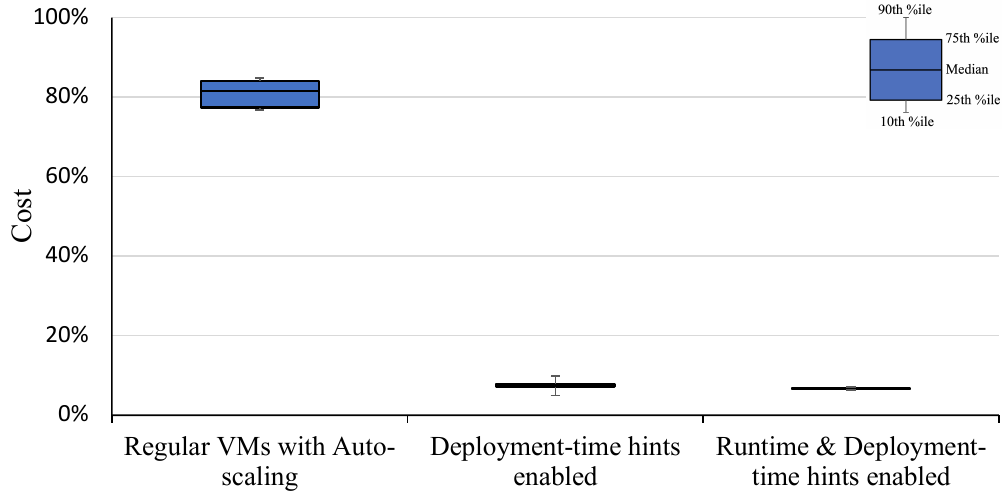}
\caption{
    Slowdown and cost running the Big Data workload.
    The baseline (\ie, 1$\times$, 100\%) is the median for Regular VMs.
}
\label{fig:hadoop-latency-cost}
\end{figure}

\myparagraph{Results}
\Cref{fig:hadoop-latency-cost} compares the baseline setup, which is Regular VMs, to WI with deployment hints and WI using both deployment and runtime hints. We also include Regular VMs with auto-scaling for comparison.
Regular VMs achieves the best performance because it constantly has access to all the resources. 
But they are also the most expensive option.
We normalize our results based on the Regular VMs performance. 

WI with deployment hints shows a median slowdown of 2.1$\times$ as it enables Auto-scaling and Harvest VMs.
However, it significantly reduces the median cost by 92.6\%.
When enabling runtime hints, the median slowdown is reduced by 21.0\%.
\edit{Based on the pricing listed in \Cref{tab:optimization-benefits},}
the cost is further reduced by 13.5\% compared to only deployment hints.

The full WI setup achieves the lowest cost (93.5\% cost reduction) while maintaining reasonable performance ($1.7\times$ slowdown for delay-tolerant workloads).
This is because in addition to leveraging Spot and Harvest VMs, the platform communicates with the workload owner to minimize the penalty of evictions and to maximize the utility of harvested resources.

\subsection{Case study: Microservices}
\label{sec:microservice}
\myparagraph{Methodology}
We deploy the social network workload from the DeathStarBench~\cite{gan2019open} on a Kubernetes~\cite{kubernetes,qiu2020firm} cluster that runs our WI extensions (\S\ref{sec:impl-workloads}).
Both control plane and worker VMs have 8 vCPUs and 32 GB of memory.
We use two management VMs and a minimum of four worker VMs.

We extend the social network~\cite{gan2019open} to mark Load Balancer, Media Frontend, Memcached, MongoDB, and Redis to run in node pools with ``management'' requirements.
The rest of the components (\ie, Nginx and the logic like post composing, timeline management, etc.) run in the node pool with ``worker'' requirements.
These worker components are load-balanced and replicated into multiple pods.
\Cref{tab:casestudy} shows the hints specified for the VMs in each node pool.

We emulate the load using a scaled-down trace from a production cluster.
We run this traffic generation from a separate set of VMs in the same virtual network.
\edit{We will also open-source this setup.}

\myparagraph{Operation}
Based on the WI hints, the platform enables oversubscription for the control VMs.
Given the load, the platform oversubscribes CPU by 50\% and memory by 20\%.

For the workers, WI enables Auto-scaling, Harvest VMs, Overclocking, and MA DCs.
\edit{
At runtime, the workload sets the runtime ``preemptibility'' hint for all the VMs except one to reduce its probability of eviction.
The workload also sets the runtime ``scale up/down'' hint to prefer harvesting and overclocking.
As Kubernetes places more containers in a worker node, it removes the ``preemptibility'' hint.
The pods with higher requirements (\eg, Redis) are deployed in a node pool that cannot be evicted.
}

In addition, MA~\cite{zhang2021flex} tracks the first set for early throttling and the second one for eviction.
In case of a power event, most components will be throttled and some workers evicted.

\myparagraph{Results}
The tail latency when running the workload in Regular VMs with autoscaling is 376 ms.
In a setup with WI where we leverage overclocking and Harvest VMs, we lower the latency down to 332 ms, which is 13.3\% improvement.
Note that we do not observe latency spikes even during evictions.

The cost for the workload owner is reduced by 44\% compared to the baseline with plain autoscaling.
Most of the savings come from running with overclocking while the rest comes from running on Harvest VMs. %

\subsection{Case study: Video conference}
\label{sec:videoconf}
\myparagraph{Methodology}
We setup a Video Conference workload on a WI-enabled cloud platform.
We extend the existing deployment scripts to provide the hints in \Cref{tab:casestudy}.
This includes deploying VMs dedicated to media processing, responsible for voice and video handling.
The load is balanced across VMs to efficiently manage calls.
The client generator is deployed separately and replicates conference traces from a production environment involving approximately 50 to 100 audio calls with 4 to 50 users, along with 2 to 75 video calls that accommodate user counts ranging from 4 to 250.
The load follows a daily pattern, with more calls during business hours.
Additionally, there are load spikes at the beginning of the hour and the half-hour mark, aligning with the start of most meetings.
\edit{For confidentiality requirements, we cannot open-source the code to run this experiment.}

\myparagraph{Operation}
The cloud platform enables various optimizations for the media-service VMs: Auto-scaling, Overclocking, Pre-provisioning, VM rightsizing, and Region-agnostic.

The local WI manager monitors the hints, while each media-service VM tracks its usage and elevates its priority during high loads.
The local overclocking controller determines whether to increase or decrease the CPU frequency of the VM based on this priority, while also factoring in the power budget and processor lifetime~\cite{jalili2021cost}.
Moreover, the right-sizing manager uses utilization data to set the right VM type for the  media-service VMs.

\myparagraph{Results}
We compare a default setup using regular VMs with the one with WI enabled. 
The WI-enabled setup is 26.3\% more cost-effective by reducing the necessary VMs for off-peak periods. %
In addition, it reduces the carbon footprint by 51\% by running VMs in a greener region.

Throughout the experiment, the workload sustains the required service level, and the conference processing rate \edit{(\ie, the number of conference calls it can handle per second)} is 35.4\% higher with WI.  %
This extra headroom indicates our conservatism, suggesting that we could have further reduced the number of VMs even further. %

WI utilizes pre-provisioned VMs during peak time to achieve two primary objectives: firstly, it effectively reduces the latency associated with adding new media-service VMs, and secondly, it enhances overall performance. This improvement is marked by a 22\% increase in conference process rates, coupled with a complete elimination of instances where conference processing encounters significant delays.

Furthermore, the WI Rightsizing Manager recommends a new VM size by monitoring utilization information from the WI Global Manager.
This rightsizing reduces costs by 13.4\%.

\subsection{Benefits at provider scale}
\label{sec:cost_analysis}

Based on the workload core usage from our survey (\Cref{tab:survey-results}), the core percentage among optimizations (\Cref{tab:cloudopt}), data from the literature~\cite{ec2-spot, azure-spot, google-spot, ambati2020slo}, internal statistics for optimizations, and our case study experimental results (\Cref{tab:optimization-benefits}), we compute the savings for workload owners when using WI. 
\edit{
To calculate the cost/carbon benefits, we need the joint probability distribution of core percentages among 10 optimizations.
Given the complexity to compute these distributions precisely, we estimate it based on the joint distribution of up to 2 optimizations and significant (\ie, >5\%) multiple-optimization scenarios via Linear Programming~\cite{linearprogramming}.
}
We derive the cost savings for each optimization and summarize them in \Cref{tab:optimization-benefits}.
We combine the results to compute the user benefit of enabling WI.
As described in \Cref{sec:coordination}, we set priorities for the optimizations that cannot be enabled simultaneously due to operational conflicts based on discount values.
Specifically, Spot VMs, Harvest VMs and Pre-provisioning do not simply provide multiplicative cost benefits when enabled at the same time due to their contention for spare compute resources.
Similarly, Overclocking, Underclocking, and MA are not compatible with each other due to CPU frequency adjustments.

\begin{figure}[t]
\centering
\includegraphics[width=0.54\columnwidth]{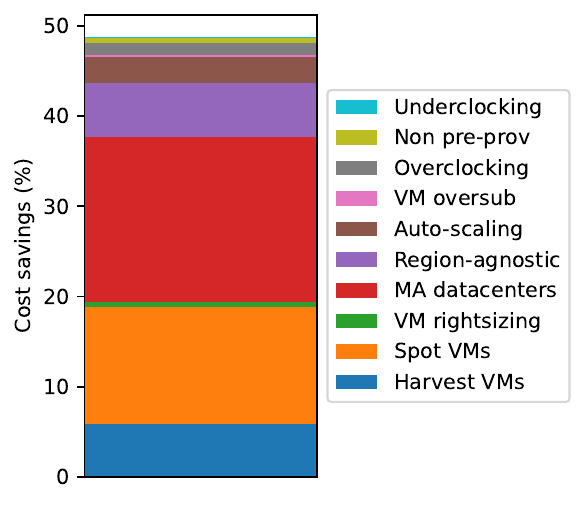}
\caption{Cost savings breakdown by optimizations.
}
\label{fig:cost}
\end{figure}

\myparagraph{Cost impact}
\Cref{fig:cost} shows the cost savings breakdown from each optimization. 
We follow the decreasing order of the owner benefits 
which mimics the workload owners' preferences when choosing their optimizations manually. 

WI reduces workload owner costs by 48.8\% on average. 
MA DCs and Spot VMs offer substantial savings by 18.3\% and 13.0\% respectively. 
Region agnostic, Harvest VMs, Auto-scaling and Overclocking also provide considerable cost reductions by 6.0\%, 5.8\%, 2.8\% and 1.3\% individually at cloud scale.
Note that actual savings vary based on owners' individual workloads, and other optimizations may play a more important role.
For example, web proxy workloads benefit from Auto-scaling, Non pre-provisioning, Overclocking, Underclocking and VM rightsizing predominantly due to its high availability requirement and the dynamic nature of web traffic.

Paradoxically, \Cref{fig:cost} shows that a higher discount from an optimization does not necessarily translate to higher cost savings. 
For example, Harvest VMs have higher discounts than Spot VMs (91\% vs. 85\%) but contribute less to the overall savings. 
This is because Harvest VMs have more strict requirements for workloads and thus result in fewer applicable scenarios.
Given workload characteristics, WI assist workload owners by automatically enabling the best set of optimizations that maximize their cost savings.

\myparagraph{Carbon impact}
We also account for the carbon generated.
Region-agnostic sends some workloads to low-carbon regions (\eg, the low 10$^{th}$ percentile) and reduce carbon footprint by 51\% (\ie, from 546 g/kWh to 267 g/kWh).
Since VM rightsizing, Auto-scaling and VM oversubscription reduce the number of VMs required to run a workload, they also reduce carbon by 50\%, 19\% and 15\% respectively.
Overall, WI can reduce carbon at provider scale by 27.6\%.

\section{Related Work}

\myparagraph{User-provided information}
Mesos~\cite{hindman2011mesos} presents a management layer that enables fine-grained resource sharing across diverse cluster computing frameworks and lets the organizations specify their policies for resource sharing. 
In high-performance computing, many works propose users giving job characteristics, such as expected job run times, which are often found to be inaccurate~\cite{tsafrir07backfilling}.
We propose to design incentives that motivate users to provide better hints.

\myparagraph{Attribute inference}
The literature on inferring attributes relevant to cloud services broadly falls into ad-hoc resource management frameworks%
~\cite{javadi2019scavenger,hadary2020protean,iqbal2022cospot} and general frameworks that combine multiple resource managers~\cite{cortez2017rc,souza2023ecovisor,wang2022sol,tang2020twine,verma2015large,tirmazi2020borg}. 
Borg~\cite{verma2015large,tirmazi2020borg} %
estimates job characteristics (\eg, resources needed, job priorities) to find suitable machines.
Twine~\cite{tang2020twine} is a framework that collaborates with applications to manage their lifecycle.
Resource Central~\cite{cortez2017rc} collects VM telemetry, learns these behaviors online, and provides predictions to various resource managers. %
SOL~\cite{wang2022sol} proposes a general on-node framework
that considers fine-grained workload dynamics, resource utilization, and provides informed decisions using ML-based agents.
To optimize carbon-efficiency, Ecovisor~\cite{souza2023ecovisor} virtualizes the energy system %
to applications.%

These works infer the attributes for various resource managers via APIs or client-side libraries.
However, all the workload hints may not be amenable for inference and these works also miss out on the relevant information from workload owners.
There still exists a gap in safe and efficient ways to leverage the resource managers as well as the inputs from workload owners for optimizing cloud efficiency.

\inigo{We should cut this a lot.}
\lexiang{I have cut a good amount.}

\myparagraph{Other cloud interfaces}
\edit{Container orchestration systems such as Kubernetes~\cite{kubernetes} and Docker Swarm~\cite{dockerswarm} automate software deployment, scaling, and management, but they are unaware of the workload characteristics running inside containers and rely on users to manually specify resource management policies.}
\edit{Serverless computing~\cite{shahrad2020serverless} provides more flexibility for the cloud platform to manage resources by allowing workload owners to upload code as tasks. However, the lack of explicit workload characteristics may lead to incorrect usage such as submitting tasks that are long-running or of low parallelism, which may result in even higher cost than using regular VMs.}
\lexiang{Added serverless for discussion.}

\section{Conclusion}

In this paper, we have proposed Workload Intelligence (WI), a novel framework for dynamic bi-directional communication between cloud workloads and cloud platform.
By punching holes through the current cloud abstraction, the platform can drastically simplify its offerings, reduce its costs without violating any workload requirements, and pass the savings to workload owners.
We have evaluated the applicability and potential of WI across 10 cloud optimizations at a major cloud provider and demonstrated significant benefits for both providers and workload owners.

\bibliographystyle{plain}
\bibliography{references}

\end{document}